# The Scheme of a Novel Methodology for Zonal Division Based on Power Transfer Distribution Factors

Michał Kłos, Karol Wawrzyniak, Marcin Jakubek, Grzegorz Oryńczak
National Centre for Nuclear Research, Świerk Computing Centre
Otwock-Świerk, Poland
karol.wawrzyniak@fuw.edu.pl

*Abstract*—One of the methodologies that carry out the division of the electrical grid into zones is based on the aggregation of nodes characterized by similar Power Transfer Distribution Factors (PTDFs). Here, we point out that satisfactory clustering algorithm should take into account two aspects. First, nodes of similar impact on cross-border lines should be grouped together. Second, cross-border power flows should be relatively insensitive to differences between real and assumed Generation Shift Key matrices. We introduce a theoretical basis of a novel clustering algorithm (BubbleClust) that fulfills these requirements and we perform a case study to illustrate social welfare consequences of the division.

*Keywords—Power system economics; Energy markets and regulation; Modelling and simulation*

## NOMENCLATURE

| | |
|---|---|
| $\mathbf{p}$ | vector of nodal injections/withdrawals |
| $\mathbf{q}$ | vector of zonal injections/withdrawals |
| $\tilde{\mathbf{p}}$ | vector of all power flow values in the system |
| $\tilde{\mathbf{r}}$ | vector of cross-border power flows |
| $\mathbf{W}$ | vector of weights denoting the importance of corresponding congestions |
| nPTDF | nodal PTDF matrix |
| zlPTDF | zonal-line PTDF matrix (the influence of zones on particular cross-border lines) |
| GSK | Generation Shift Keys matrix |
| $N$ | number of nodes in the system |
| $M$ | number of lines in the system |
| $L$ | number of cross-border lines |
| $K$ | number of congested cross-border lines |
| $pre/act$ | indices denoting predicted/actual values |

## I. INTRODUCTION

The energy market in Europe is undergoing a process of transformation aimed at integration of national markets and making better use of renewable generation sources. The market structure used in many countries, mostly due to historical reasons, is the uniform pricing, in which there is a single price of energy set on a national market for each hour of a day. In spite of its apparent simplicity, such an approach has serious disadvantages. The equilibrium set on the market does not take into account safety requirements of the grid. Hence, (i) the single-price equilibrium set on the market (energy exchange) is frequently unfeasible, (ii) the system operator has to perform costly readjustments, (iii) costs of supplying the energy differ between locations, but they are not covered where they arise.[1] With introduction of other forms of market, the congestion costs are mitigated and the price on the market reflects the true costs of supplying energy to different locations in a more adequate way.

Wholesale electricity markets use different market designs to handle congestion in the transmission network. The two most popular approaches towards which national markets evolve are nodal and zonal pricing. The nodal pricing model is currently used in, among others, the US and Russia. Zonal pricing has been introduced in the Nordic countries as well as in Great Britain. Currently European Network of Transmission System Operators for Electricity (ENTSO-E) is under the process of introducing zonal model as the default model for the whole continent. The algorithm responsible for determining zonal configuration is the main subject of this paper.

Zonal market, which can be thought of as a compromise between simplicity of uniform structure and accuracy of nodal one, introduces differentiation of prices between regions with distinct costs of supplying energy. The power grid is divided into geographical regions (zones), each having a separate market for the energy with possibly different price. Market Coupling (MC) algorithm is used to control inter-zonal power flows and to calculate prices in zones given those flows. This way, under presumption that the zones were chosen so that frequently congested lines are on their borders, the equilibrium on zonal markets will take into account the transfer limits on those critical lines. The need for additional congestion management is thus minimized, with most of the task being performed by the MC mechanisms. Of course, small adjustments of equilibrium to satisfy limits on intra-zonal lines

---

[1] For example, in Poland in 2011 the cost of the balancing market readjustments amounted to more than 3% (>250 Mln EUR) of the overall costs of production (*source: URE/ARE S.A.*).

This work was supported by the EU and MSHE grant nr POIG.02.03.00-00-013/09.

might be necessary, but they are expected to be less costly than the adjustments on a corresponding uniform market.

Still, there is no consensus in the literature with respect to methodology of identification of optimal zones' number and their borders. The existing methods are mostly based on two-stage approach – assignment of some specific values to each of the nodes and division of the system into regions by clustering the nodes over those parameters. Among prominent methods, we can distinguish two popular approaches for choosing the values characterizing nodes.

The first approach is based on *nodal prices* called also Locational Marginal Prices (LMPs). Vast literature ([1]-[4], among others) covers various attempts to utilize this method. Nodal price represents the local value of energy, i.e. a cost of supplying extra 1 MW of energy to a particular node - physical point in the transmission system, where the energy can be injected by generators or withdrawn by loads. This price consists of the cost of producing energy used at a given node and the cost of delivering it there taking into account congestion. Therefore LMPs separate locations into higher and lower price areas if congestion occurs between them. The nodes/locations with similar prices are then grouped (clustered) to determine the candidates for market zones.

The second approach is based on Power Transfer Distribution Factors (PTDFs). The procedure starts from identification of congested lines, for which the PTDFs are then calculated. The distribution factors reflect the influence of unit nodal injections on power flow along the transmission lines, thus grouping the nodes characterized by similar factors into one zone defines a region of desirably similar sensitivity to congestions. The two notable variants of PTDF approach proposed in the literature are [5] and [6].

In the first of them [5], the nodes which have the highest absolute values of PTDFs with respect to congested line are placed in one zone, while nodes which have similar and small in magnitude PTDFs in other zones. Consequently, the characteristic of PTDFs usually place the congested line in the middle of the zone with variable PTDFs, thus this approach is not in line with the postulate that the congestion should be managed as much as possible by the mechanisms of the inter-zonal MC.

Approach [6] makes use of the PTDF matrix in order to obtain a clear-cut distinction whether a power injection in a particular node increases the flow (in a given direction) over a line, or decreases it. Then a partition of the grid is made step by step according to subsequent congested lines.

The other method utilizing distribution factors [7] is based on k-means algorithm and aims at grouping nodes by similar values of PTDFs. The limitations of k-means clustering are however widely discussed [8] and concern mainly the problem of finding the optimal initialization (seed).

In this paper we introduce the new concept of grouping nodes in multidimensional space instead of grouping only along one dimension at once [6]. Namely, we consider the influence of nodes on all critical branches by which we identify groups of nodes characterized by similar impact on power flow increase. Multiple advantages of the method are discussed in comparison to outcomes of other algorithms (cf. sec. III-IV).

The exact methodology is introduced in Sec. II. In Sec. III we present case study to assess social welfare under divisions obtained from ours and comparable methods ([5],[6]), while in Sec. IV we conclude and point out issues which would be necessary to make the methodology more accurate.

## II. THE METHODOLOGY

In Subsection A we will show the objective function that we minimize. Minimization requires identification of congested lines (Subsec. B) and proper method of nodes' aggregation (Subsec. C). In our case the aggregation is conducted in so called PTDF space (Subsec. D). The complete algorithm is provided in Subsec. E.

### A. Objective Function

One of the possible objectives for defining optimal partition is to aim at maximizing social welfare (social welfare is defined as a sum of consumer and producer surpluses for each bidding area [10]). This criterion is often chosen as a universal measure of optimality for all market structures (uniform, zonal and nodal) and may be used in order to compare them in an objective manner. Another criterion is based on the accuracy of power flows prediction obtained from MC algorithm and safety measures. It reduces firmness costs (the cost of remedial actions) and maximizes allowed power flow between zones. We provide a more detailed explanation below.

The zonal energy market can be represented as a set of energy exchanges, each governing the trade between generators and consumers[2] of energy located in the particular geographic area. Energy transfers between the zones are allowed and are governed by MC mechanisms, which take into account the capacity constraints on the inter-zonal transmission lines. In order to determine safe supply-demand equilibria on each of energy exchanges, the MC mechanism must determine how the realization of buy/sell bids translate into (i) power injections/withdrawals in the nodes of the grids and into (ii) flows on inter-zonal lines. The issue of great importance for MC is an a priori assumption about the levels of load and generation expressed by so called Generation Shift Keys matrix (GSK). This constitutes serious paradox as these levels are to be determined as an output of MC. Hence, some prediction of the output has to be also assumed as an input. If it is wrongly guessed, then the solution indicated by MC can be potentially very far from reality. This paradox is unsolvable on the level of MC implementation, the only solution is to provide zonal divisions which are insensitive to the GSK deviations. Further we will show how to choose nodes in a way that reduces impact of misestimated GSK matrix.

MC procedures acting on day-ahead basis must in consequence work on a prediction of generation/load pattern. By $\mathbf{p} = (p_1,...,p_N)$ we denote the vector[3] of assumed power injection/withdrawal in all the $N$ nodes in the grid. Inaccuracy of the day-ahead prediction of those injections/withdrawals ($\mathbf{p}^{pre} \neq \mathbf{p}^{act}$, where $\mathbf{p}^{pre}$ denotes predicted injections, and $\mathbf{p}^{act}$ actual injections) leads to miscalculation of power flows, which are denoted by $\tilde{\mathbf{p}} = (\tilde{p}_1, ..., \tilde{p}_M)$, on all the $M$ lines of the grid, particularly on the inter-zonal lines which are controlled by MC process. We denote by $\tilde{\mathbf{r}} = (\tilde{r}_1, ..., \tilde{r}_L)$ the subset of $(\tilde{p}_1, ..., \tilde{p}_M)$ which represents the $L$ inter-zonal lines.[4]

---
[2] By "consumers of energy" we do not mean the single households/enterprises, but the distribution companies buying the energy on wholesome market to satisfy the demand in their area.
[3] In all following matrix equations vectors have to be interpreted as columns. The transposition symbol (T) is omitted in all definitions.

The miscalculation of power transfer on inter-zonal lines during day-ahead MC process can be of two types. Underestimation ($\tilde{r}_l^{pre} < \tilde{r}_l^{act}$) can lead to physical (thermal) damage to transmission line and threat of outages due to violation of maximal power limit. Overestimating the exchanged power ($\tilde{r}_l^{pre} > \tilde{r}_l^{act}$) forces MC algorithm to restrain the exchange of the permitted amount of power between markets, which results in economic inefficiency.

Thus, the accuracy of prediction of power flows is crucial both for the safety of the transmission network, as well as for the economic efficiency of zonal market. Additionally, the choice of inter-zonal lines (lines lying on zones' borders) determines how much of the possible congestion will be controlled by the mechanisms of MC, and how much by the intra-zonal congestion management. Since the congestion management generates hidden costs on the energy market, our aim of both economic efficiency and safety of the grid is approximated by the following objectives of the zonal division:

(i) find the frequently-congested lines in operating conditions of the power market and place them on the zones' borders, so that their congestion will be managed in a transparent manner by MC mechanisms;

(ii) cluster the nodes in such a manner that minimizes the prediction error of the inter-zonal flows' forecast on frequently congesting lines, that is $\min \|\Delta \tilde{\mathbf{r}}\|$, where $\|.\|$ denotes the Euclidean vector norm and $\Delta \tilde{\mathbf{r}} \equiv \tilde{\mathbf{r}}^{pre} - \tilde{\mathbf{r}}^{act}$.

### B. Congestion Identification

The first step of the proposed procedure aimed at an electrical grid's partitioning is the identification of frequently congested lines in operating conditions of the power market and the grid. There are many methods used for identification. E.g. it can be conducted based on experts' knowledge [9], by CCI [6] or by Optimal Power Flow (OPF) algorithm, which is run to determine the least-cost conventional generation scheme. The Karush-Kuhn-Tucker (KKT) multiplier assigned to every line describes the added cost of shifting generation necessary to alleviate the congestion on this line. The lines with the highest average congestion costs $k_i$ (we have different runs for different load and generation scenarios) are then the natural "candidates" for the inter-zonal lines, over which the transfers should be then controlled by MC mechanisms. Such approach is presumed to best serve the aim of minimizing the costs of intra-zonal congestion management necessary after reaching equilibrium on a zonal market.

Next, we construct vector of weights $\mathbf{W} = (W_1, ..., W_L)$ representing the average congestion cost on line $j$ scaled by the sum of average congestion costs:

$$W_j = \frac{k_j}{\sum_{i=1,...,k} k_i}. \quad (1)$$

Our procedure differs substantially from the one employed in [6]. There, starting with a market equilibrium with no transfer limits taken into account, three types of steps are undertaken sequentially: (i) identification of the most congested line (in terms of percentage of overload in equilibrium when the limit is not taken into account), (ii) division of the grid with respect to this line, (iii) addition of the transfer limit of this line into the equilibrium constraints. In such sequential approach, the actual congestion cost of a line is not easily derivable – the market equilibrium changes from one iteration to another, as well as the power flow across transmission lines in the grid. Our approach, making use of the characteristics of KKT multipliers, derive the actual cost which a particular line adds to the system in the point of equilibrium, which takes into account all the transfer limits at once. Additionally, our approach results in a vector of weights representing the relative magnitude of the congestion costs of the transmission lines – an object which will be of importance in the next steps of the division procedure.

### C. Linear Operators

There are many variants of Power Transfer Distribution Factors (PTDF) matrices. For further considerations we need to introduce two of them. By nPTDF we denote the nodal matrix of $M$ (lines) by $N$ (nodes) elements. The matrix is used to transform nodal injections $\mathbf{p}$ into power flows $\tilde{\mathbf{p}}$ in the following way:

$$\tilde{\mathbf{p}} = \text{nPTDF} \, \mathbf{p}.$$

Another operator, zonal variant of PTDF matrix – zlPTDF ('zonal-line[5] PTDF' [11]), can be applied if net positions of all the bidding areas are known, which means that we need to construct a vector of *zonal* injections $\mathbf{q} = (q_1, ..., q_Z)$. Particular coordinates $q_j$ are calculated as

$$q_j = \sum_{\text{node } i \in \text{zone } j} p_i.$$

*Zonal-line* PTDF is a product of nodal PTDF and Generation Shift Keys (GSK). GSK expresses the influence of a node in a zone on the net export of the zone. The GSK is an $N$ (nodes) by $Z$ (zones) matrix. Each element $GSK_{ij}$ is equal to zero if node $i$ is not included into zone $j$, otherwise it is equal to the ratio of nodal injection $p_i$ to zonal net position $q_j$. Thus, GSK depends on the generation pattern $\mathbf{p}$ and zonal attribution of nodes, and we have that

$$\tilde{\mathbf{p}} = \text{zlPTDF} \, \mathbf{q} = \text{nPTDF} \, \text{GSK} \, \mathbf{q}.$$

---

[4] The limits of intra-zonal lines are assumed to be governed by zonal congestion management. In the further part of the article we will argue that our proposed methodology leads to a zonal partition in which the intra-zonal congestion should not be a frequent issue.

[5] We use 'zonal-line' term for PTDF matrices that transform zonal injections into power flows on existing transmission lines – which is worth mentioning, as the other popular *zonal* (or 'zonal-interface' [10]) PTDF works as operator transforming zonal net positions to *cumulative* inter-zonal power exchanges (the sum of all power transfers across each of the borders).

By $\tilde{\mathbf{p}}^{pre}$ we denote the power flow resulting from *actual* generation pattern, but *predicted* (and, in consequence, possibly outdated) GSK. This configuration of factors forms inevitable miscalculation which is crucial for robustness of inter-zonal MC mechanism. Predicted GSK becomes known to all market participants as Transmission System Operator (TSO) is obliged to publish it before the bids are submitted at the energy exchange. The same, anticipated form of GSK is being used as an input for MC algorithm, which determines whether a bid is accepted or rejected. As the result of the bids' acceptance procedure a new, actual pattern of nodal injections is found ($\mathbf{p}^{act}$). Hence the MC equilibrium is based on predicted GSK and actual $\mathbf{p}$. On the other hand, real generation/load scenario can be used to determine actual power flows transmitted via all the grid's branches, $\tilde{\mathbf{p}}^{act}$. The differences between predicted and actual (real) power flows are given by

$$\begin{aligned}\tilde{\mathbf{p}}^{pre} - \tilde{\mathbf{p}}^{act} &= \text{zlPTDF}^{pre}\,\mathbf{q}^{act} - \text{zlPTDF}^{act}\,\mathbf{q}^{act} \\ &= \text{nPTDF}(\text{GSK}^{pre} - \text{GSK}^{act})\mathbf{q}^{act} \quad (2) \\ &= \text{nPTDF}\,\Delta\text{GSK}\,\mathbf{q} = \Delta\tilde{\mathbf{p}}.\end{aligned}$$

Equation (2) shows that there are three factors determining the error of power flow vector estimator: magnitude of nPTDF, the accuracy of prediction of GSK and the magnitude of zonal injections. In what follows, we will concentrate on the first of those factors.

We note that the sum of elements in each column of GSK matrix equals unity, since each of the coefficients can be interpreted as a percentage of nodal contribution to overall zonal net position (export or import). Let us assume that $\text{GSK}^{x}$, $\text{GSK}^{y}$ are some instances of GSK. Then $\forall j\ \sum_i \text{GSK}^x_{ij} = \sum_i \text{GSK}^y_{ij} = 1$. Thus, the difference of any two GSKs is an operator, in which the elements along each column sum up to zero:

$$\forall x, y\ \forall j\ \sum_i \left(\text{GSK}^x_{ij} - \text{GSK}^y_{ij}\right) = \sum_i \Delta\text{GSK}_{ij} = 0.$$

From (2) we get that the power flow error on $k$-th transmission line can be expressed by the following sum:

$$\Delta\tilde{p}_k = \sum_j \underbrace{\left(\sum_i \text{nPTDF}_{ki}\Delta\text{GSK}_{ij}\right)}_{\Delta\text{zlPTDF}_{kj}} q_j.$$

As the columns of GSK along index $i$ sum up to zero, the value of whole product ($\Delta\text{zlPTDF}_{kj}$) will be zero if the elements of $i$-th row nPTDF are identical.

Obviously, we cannot expect the rows of nPTDF to consist of equal-valued elements. Nevertheless, minimizing the differences between the nodal PTDFs inside a zone or, equivalently, the intra-zone variance of nodal PTDFs, will lead to a smaller error of power flow's prediction along the line $k$.

*D. PTDF Space*

The previous section concludes with description of procedure to minimize the flow prediction error for a single transmission line. Facing the need for minimization of the prediction error along multiple frequently congested lines $\tilde{\mathbf{r}}$, we present below a multidimensional formalism, which will allow for nPTDF-variance-minimizing clustering with respect to more than one line at the time.

The main postulates of PTDF space come down to the following assumptions:

(i) the columns of nPTDF are treated as vectors in $M$-dimensional space. Each vector corresponds to a certain node and its coordinates denote the amount of power transmitted through a certain line as a fraction of a power unit injected in this node. This unambiguous measure of power flow is possible, given that one reference node is chosen as a "sink," which is responsible for withdrawing all the power transmitted through the system. In the appendix we show that the choice of the reference node does not affect in any way the results of the clustering with respect to nPTDFs;

(ii) since not all of the lines demand as much attention due to a different degree of congestion, we postulate taking into account the relative congestion degree by scaling all the vector's coordinates by congestion rate factors $\mathbf{W}$, defined in (1). This method enhances the role of the lines which add the most congestion costs into the system and diminishes the importance of lines which are rarely congested or their congestion is relatively cheap to manage.

As the result, each vector is represented by one point in $M$-dimensional space. Similarity of nodes may be examined by comparing their distances using any proper metric (e.g. Euclidean, Manhattan, etc.). Three important properties characterize the PTDF space:

(a) nodes lying on the ends of a congested line are far from each other. If we consider any two nodes situated at the verges of a congested line $l$, their coordinates in PTDF space corresponding to $l$-th line will constitute extreme values in this vector. This undeniably useful property prevents from grouping such two nodes into one zone, which follows the objective (i) sec. II A;

(b) the number of significant dimensions is strictly related to weights $W_l$ and/or the threshold defined arbitrarily by the examiner. Null or small values of $\mathbf{W}$ indicate the dimensions which do not play any role in defining the position in PTDF space, thus it is convenient to exclude them from the analysis. For instance, if $\tilde{\mathbf{r}}$ is a 10-dimensional vector of power flows and among them we observe only 2 significant congestions, the PTDF space for the problem will be reduced to 2D plane. Moreover, contrary to [6], the method can treat multiple lines congested similarly frequently in a proportionally similar way, preventing significant qualitative asymmetry if the rates of congestion differs only minutely (which can result i.e. from statistical fluctuations);

(c) the choice of reference node leaves the analysis unaffected – changing reference node is equivalent to simultaneous translation of all the points by the same vector, which obviously does not influence distances between any pair of them (cf. Appendix).

*E. Clustering Algorithm*

The PTDF space-based algorithm designed to achieve all the above objectives takes the following data as an input:

(i) the nPTDF matrix for the grid,

(ii) congestion rates matrix, which consists of two columns; first, the numbers of nPTDF's rows corresponding to frequently congested lines $\tilde{r}$, second, the weight factors reflecting congestion costs for these branches ($\mathbf{W}$).

The congestion rates matrix is used to determine a subspace of the full *M*-dimensional PTDF space by selecting only the products of coordinates and corresponding $W_l$ coefficients which contribute with nonzero outcomes to the error on congested lines, $\Delta\tilde{r}$. The nodes are then represented by a set of coordinates with respect to the *K* mostly congested lines (all the lines with corresponding $W_l > 0$). The algorithm, named "BubbleClust," performs then the zonal division in two stages:

(I) we start with $k$ ($k \leq 2K$, inequality takes place when some node is connected to two or more congested lines) initial singleton zones which are equivalent to the verges of previously identified congested lines. Each zone is characterized by the coordinates of its center. In every step we evaluate the Euclidean distance between the centers and nodes adjacent (topologically) to zones. Next, the node characterized by minimal distance is included into the proper zone forcing recalculation of its center and update of the set of 'free' adjacent nodes. The process continues until each node is assigned to some zone;

(II) having obtained a pre-partition into $k$ zones, in each step, the algorithm continues to merge two of the closest (in the Euclidean distance between zones' centers) adjacent (in grid topology) areas;

Finally, the algorithm outputs a set of possible division into any number of zones $z$, such, that $2 \leq z \leq k$ (cf. Fig 1).

### III. CASE STUDY

As both cited studies concerning PTDF-based zonal division present application of their algorithms on the example of New England IEEE 39 bus system, we also follow this custom for the sake of comparability. We modify the case using the guidelines given by [6], [12] (nodal consumption, generation cost and flow capability). Using this data and BubbleClust algorithm we obtained two divisions: the one which does not allow zone without generators (ver. 1), the second – less rigorous, without the aforementioned restriction (ver. 2).

We evaluated four different scenarios of division into four zones: Kumar's (recreating division of [5]), Kang's (recreating [6]) and two PTDF space-based, using one of the most commonly accepted measures – social welfare (SW), which is the sum of consumer surplus, producer surplus and congestion rent. The exact attribution of nodes of IEEE 39 bus system is presented in the appendix. The strict definition of parameters and detailed discussion of the method can be found in [13]. Computation of social welfare was possible due to implementation of Market Coupling algorithm, the method used in CWE (Central Western European) region since 2011. The results of the comparison are illustrated in Tab. 1.

The analysis of the results leads to a few conclusions. Firstly, all the outputs show some similarities. We can distinguish between three nodal groups that constitute separate zones in all cases (cf. Tab. 2). Relatively small group 3 in three out of four cases could be extended by nodes 19, 20, 23, 24, 33, 34, 36.

The aforementioned resemblance between different zonal configurations is the effect of structural features of the system common in all the cases, i.e. values of line susceptances. Congestion Contribution Identification process (CCI) [6] finds three lines: 16-24, 3-4 and 16-19.

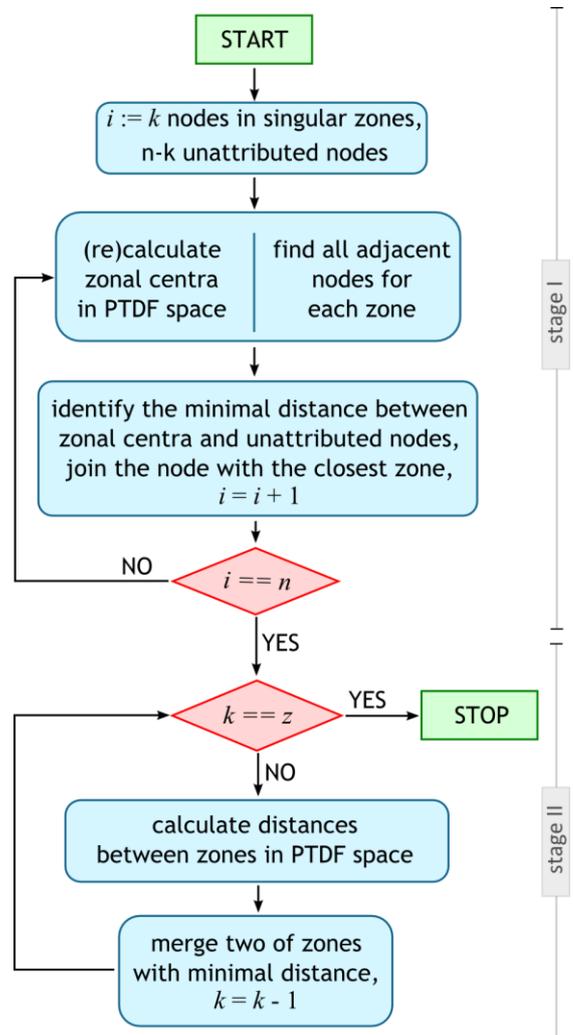

Figure 1. Flow chart of BubbleClust algorithm.

| Alg. | Kumar | Kang | BubbleClust v.1 | BubbleClust v.2 |
|---|---|---|---|---|
| SW | 0 | 16 | 326 | 830 |

Table 1. The difference between social welfare calculated for the four clustering methods and SW of division performed by Kumar's algorithm. The values are expressed in USD.

| Group 1 | 4, 5, 6, 10, 11, 12, 13, 14, 31, 32 |
|---|---|
| Group 2 | 16, 21, 22, 35 |
| Group 3 | 26, 28, 29, 38 |

Table 2. Three groups of nodes which are included into separate zones in all analyzed cases.

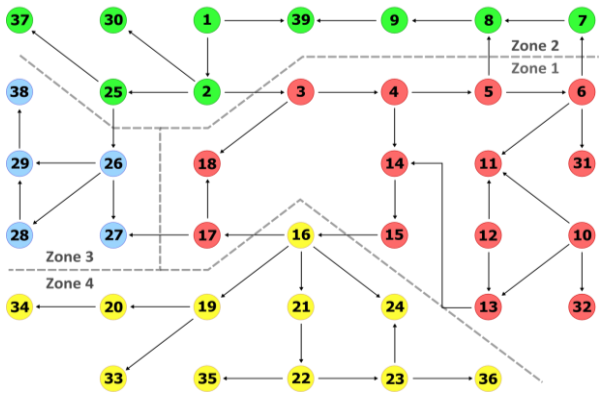
algorithm of A. Kumar *et. al.* [5]

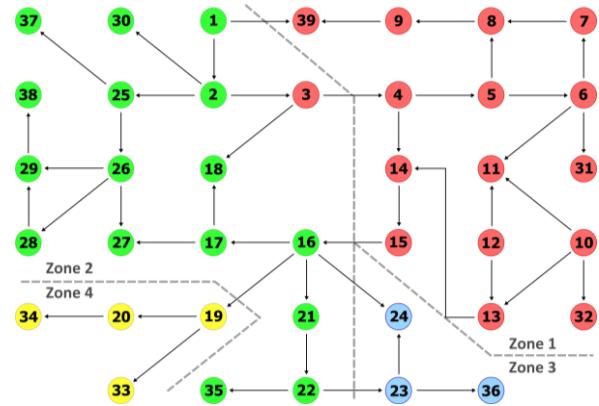
algorithm of C.Q. Kang *et al.* [6]

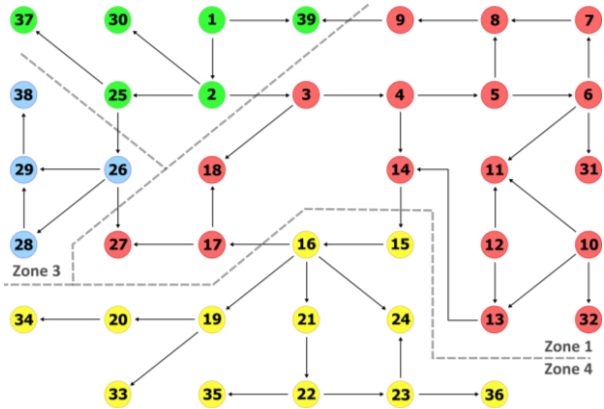
BubbleClust in PTDF space (ver. 1)

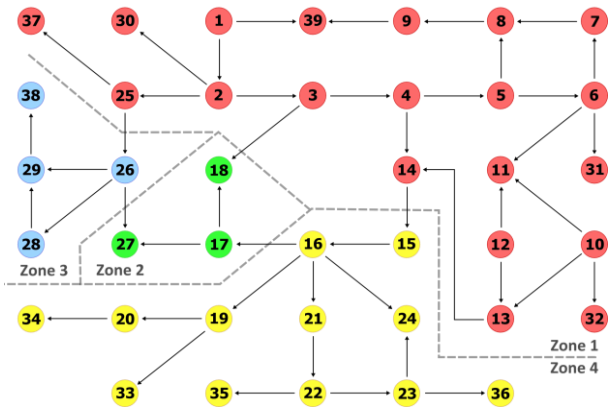
BubbleClust in PTDF space (ver. 2)

Figure 2. The partition of IEEE 39 bus system of New England; four different methods result in various realization of zonal division.

## IV. CONCLUSIONS & FUTURE WORK

The improvement obtained by Kang's solution over the one given by Kumar [5] can be explained by introducing more realistic economic impact; CCI discovers bottlenecks one by one by systematic constraining of initially unlimited transmission capabilities. This, however, does not change the fact that CCI bases on artificial heuristics by starting with infinite transmission capacity characterizing all connections in the system. As an input for BubbleClust algorithm, we use four most congested lines resulting from Optimal Power Flow based dispatch (the overloads take place on the following lines: 16-24, 26-27, 17-18, 25-26 and 3-18), the objective functions of OPF and Market Coupling are not identical, but in case of inelastic demand, the resemblance[6] is good enough to increase the result of the assessment significantly.

The difference between two PTDF space-based results is caused by neglecting the least important congestion (3-18) in order to satisfy the condition of including generation into each zone. Ignoring this assumption (case "BubbleClust ver. 2") leads to incorporation of two generators (30 & 37) into the biggest zone and delimitation of small bidding area (17, 18, 27) with no direct connection with any source of power (which does not affect the zonal demand as long as the transmission capability allows the adequate amount of imported power).

Although we managed to prove the usefulness of the developed method, we ought to make two important remarks. The estimation of social welfare would be more accurate if redispatching costs were included. Taking operational readjustments into the account will be the domain of further research. Second, the overall social gain resulting from choosing the division method is rather small if juxtaposed with 6.3 GW of total load and social welfare of the magnitude of millions of dollars in each scenario. Nevertheless, 39 nodes constitute a space which is not complex enough to illustrate all the potential benefits. Real-world problems involve thousands of buses connected by thousands of transmission lines. Applying discussed methods to large-scale cases could lead to more spectacular differences in the evaluation stage.

We presented a methodology of zonal division of the energy market, which aims to satisfy both economic (control inter-zonal congestion in a transparent manner, minimize the added costs of intra-zone congestion management) and system stability (accuracy of prediction of flows on the critical, frequently congested lines) criteria. We based our method on clustering in multi-dimensional nPTDF space of coefficients related to flows over the congested lines. This presentation, however, should be treated as preliminary, since there is still couple of issues being currently worked on.

---

[6] Social welfare maximized by Market Coupling is similar to minimization of generation costs done by OPF. The difference concerns the possibility of unsatisfied demand and does not play a big role in such small system.

First, we use extensively in our approach the GSK matrices to translate zonal injections into power flows. Since GSK consists of the ratios of nodal injections to zonal net positions, it is meaningless ("singular") if there is any zone that is completely self-sufficient i. e. its net position equals zero. In such situation the GSK matrix - as well as the calculation of flows on basis of it - is undefined, and our methodology does not produce any meaningful results. Also, when the net position of a zone is close to zero, the flow calculation is not numerically stable. We are working currently on an alternative specification of the GSK matrix, which would overcome the aforementioned problems.

Additionally, we treat symmetrically over- and underestimation of power flows on the critical lines in our error measure. Still, a prediction which overestimates the exchanged power is "safer" for the system, since it leads only to an economic loss in market efficiency due to more constrained MC mechanism. Underestimation, in turn, may lead to a physical damage of the power lines, power outage and blackout in a significant part of the grid. Thus, prediction error which underestimates the power flows should be penalized much more than the one overestimating them. This would require a non-linear specification of the error terms similar to that of $\Delta \tilde{\mathbf{r}}$. The specification of an asymmetric error measure is also a work in progress.

## APPENDIX

We now prove that the choice of the reference node does not affect any calculations on PTDF matrices significant in our division methodology. Let us assume that $\text{PTDF}^i \in \mathbb{R}^{M \times N}$ is a nodal PTDF matrix constructed under an assumption that $i$ is the reference node. The decision which node we use as the reference one is crucial when the analysis of separate matrices' elements is concerned, but is irrelevant as long as we use $\text{PTDF}^i$ matrix only for calculating power flows. In fact, all $\text{PTDF}^i$, $i \in \{1, ..., N\}$, constitute an equivalence class with respect to left-handed multiplication by vectors $\mathbf{p} = (p_1, ..., p_N)$ such that $\sum_{n=1}^{N} p_n = 0$ (generation/load balance in lossless transmission). In other words, for

$$\tilde{\mathbf{p}} = \text{PTDF}^1 \mathbf{p} = \text{PTDF}^2 \mathbf{p} = \ldots = \text{PTDF}^N \mathbf{p},$$

the product $\tilde{\mathbf{p}}$ remains unaffected if we apply any of $\text{PTDF}^i$ operators to any vector $\mathbf{p}$, which, due to a lossless network equilibrium constraints, obviously satisfies the aforementioned network balance property. In other words, clusterings in such PTDF spaces will lead to the same result, since distances between points remain unchanged by the shift of origin.

Let us prove that adding the same element $\alpha_l$ to selected row of matrix $\text{PTDF}^i$ does not influence the left-hand side:

$$\tilde{p}_l = \sum_{n=1}^{N} (\text{PTDF}^i_{ln} + \alpha_l) p_n = \sum_{n=1}^{N} \text{PTDF}^i_{ln} p_n + \alpha_l \underbrace{\sum_{n=1}^{N} p_n}_{0}$$
$$= \sum_{n=1}^{N} \text{PTDF}^i_{ln} p_n.$$

In the consequence, we may choose the coordinates of vector $\boldsymbol{\alpha}$ and create new PTDF operator $\mathbf{S} = \text{PTDF}^i + \boldsymbol{\alpha} \mathbf{u}^T$, as the sum of PTDF matrix and a dyadic product of $\boldsymbol{\alpha}$ and $N$-dimensional vector $\mathbf{u} = (1, 1, ..., 1)$. $\mathbf{S}$ inherits all operational properties of nPTDF matrix, but, on the other hand, different explicit instances of this operator lead to different coordinates of points in PTDF space specifically, to a translation of the points' coordinates by $\boldsymbol{\alpha}$, thus causing no impact on the mutual distances between those points.

## REFERENCES

[1] B. Burstedde, "From Nodal to Zonal Pricing – A Bottom-Up Approach to the Second-Best," Proceedings of the 9th EEM, May 2012, pp. 885-892.

[2] J. Bialek and M. Imran, "Effectiveness of Zonal Congestion Management in the European Electricity Market," Proceedings of the PES PowerAfrica, July 2007.

[3] C. Breuer et al., "Determination of Alternative Bidding Areas Based on a Full Nodal Pricing Approach" Power and Energy Society General Meeting (PES), 2013 IEEE.

[4] C. Breuer, A. Moser, "Optimized Bidding Area Delimitation and Their Impact on Electricity Markets and Congestion Management" Conference and Proceedings: European Energy Markets 2014.

[5] A. Kumar et al., "A Zonal Congestion Management Approach Using Real and Reactive Power Rescheduling," IEEE Transactions on Power Systems, vol. 19, no. 1, Feb. 2004, pp. 554-562.

[6] C.Q. Kang et al., "Zonal marginal pricing approach based on sequential notwork partition and congestion contribution identification," Electrical Power and Energy Systems, vol. 51, Oct. 2013, pp. 321-328.

[7] C. L. Duthaler, "A Network- and Performance-based Zonal Configuration Algorithm for Electricity Systems", PhD dissertation, EPFL, Lozanna 2012.

[8] F. Wang, K. W. Hedman, "Reserve Zone Determination Based on Statistical Clustering Methods", North American Power Symposium (NAPS), 2012.

[9] ENTSO-E, "Technical Report: Bidding Zone Review Process", 2013.

[10] K. Wawrzyniak et al., "Division of the Energy Market into Zones in Variable Weather Conditions using Locational Marginal Prices," forthcoming in Proceedings of the 39th IECON 2013 [http://arxiv-web3.library.cornell.edu/pdf/1310.5022].

[11] J-B. Bart and M. Andreewsky, "Network Modelling for Congestion Management: Zonal Representation Versus Nodal Representation," Proceedings of the 15th PSCC, August 2005.

[12] S.H. Kim, J.U. Lim and S.I. Moon, "Enhancement of Power System Security Level through The Power Flow Control of UPFC," IEEE PES Summer Meeting, vol. 1, Seattle, 2000, pp. 38–43.

[13] G. Oryńczak, M. Jakubek, K. Wawrzyniak and M. Kłos, "Market Coupling as the Universal Algorithm to Assess Zonal Divisions," Conference and Proceedings: European Energy Markets 2014.